# Quantum Cryptography

Richard J. Hughes

D. M. Alde, P. Dyer, G. G. Luther, G. L. Morgan and M. Schauer

*University of California*
*Physics Division*
*Los Alamos National Laboratory*
*Los Alamos, NM 87545*

## ABSTRACT

Quantum cryptography is a new method for secret communications offering the ultimate security assurance of the inviolability of a Law of Nature. In this paper we shall describe the theory of quantum cryptography, its potential relevance and the development of a prototype system at Los Alamos, which utilises the phenomenon of single-photon interference to perform quantum cryptography over an optical fiber communications link.

"I am fairly familiar with all forms of secret writings, and am myself the author of a trifling monograph upon the subject, in which I analyse one hundred and sixty separate ciphers; but I confess that this one is entirely new to me. The object of those who invented the system has apparently been to conceal that these characters convey a message ..." (Sherlock Holmes.[1])

## 1. Introduction

Cryptology, the mathematical science of secret communications, has a long and distinguished history of military and diplomatic uses dating back to the ancient Greeks.[2] In World War II, Allied successes in breaking the ciphers of Germany and Japan played an important part in the outcome of the conflict and the development of the modern computer.[3] Today, the ability to ensure the secrecy of military or diplomatic communications is as vital as ever, but cryptography is also becoming more and more important in everyday life. With the growth of computer networks for business transactions and communication of confidential information there is an ever increasing need for encryption to ensure that this information cannot be acquired by third parties. Remarkably, the seemingly unrelated philosophical foundations of quantum mechanics are now being brought to bear directly on the problem of communications security in the potentially practical emerging technology of quantum cryptography.

In this paper we shall review the theory of quantum cryptography, its potential applications and the development of an experimental prototype at Los Alamos. We shall answer the questions: What is "quantum" about quantum cryptography? Will we need it? or, what if anything is "wrong" with conventional cryptography? What are the limitations imposed by "practical" issues such as losses and noise? What are the prospects for future improvements? What hardware developments are desirable? and, What kind of secure communications problem could it be used for?

The two main goals of cryptography are for a sender and an intended recipient to be able to communicate in a form that is unintelligible to third parties, and for the authentication of messages to prove that they were not altered in transit. Both of these goals can be accomplished with provable security if sender and recipient are in possession of shared, secret "key" material. Thus, key material, which is a truly random number sequence, is a very valuable commodity even though it conveys no useful information itself. One of the principal problems of cryptography is therefore the so-called "key distribution problem." How do the sender and intended recipient come into possession of secret key material while being sure that third parties ("eavesdroppers") cannot acquire even partial information about it? It is provably impossible to establish a secret key with conventional communications, and so key distribution has relied on the establishment of a physically secure channel ("trusted couriers") or the conditional security of "difficult" mathematical problems in public key cryptography. However, provably secure key distribution becomes possible with quantum communications. It is this procedure of key distribution that is accomplished by quantum cryptography, and not the transmission of an encrypted message itself. Hence, a more accurate name is quantum key distribution (QKD).

The most obvious security feature of QKD is that it is impossible to "tap" single quantum signals in the conventional sense. At a deeper level, QKD resists interception and retransmission by an eavesdropper because in quantum mechanics, in contrast to the classical world, the result of a measurement cannot be thought of as revealing a "possessed value" of a quantum state. Moreover, Heisenberg's uncertainty principle ensures that the eavesdropper's activities must produce an irreversible change in the quantum states ("collapse of the wavefunction") before they are retransmitted to the intended recipient. These changes will introduce an anomalously high error rate in the transmissions between the sender and intended recipient, allowing them to detect the attempted eavesdropping. Thus, the two important security features of QKD are that eavesdroppers cannot reliably acquire key material, and any attempt to do so will be detectable.

The origins of quantum cryptography can be traced to the work of Wiesner, who proposed that if single-quantum states could be stored for long periods of time they could be used as counterfeit-proof money. Wiesner eventually published his ideas in 1983,[4] but they were of largely academic interest owing to the impracticality of isolating a quantum state from the environment for long time periods. However, Bennett and Brassard realised that instead of using single quanta for information storage they could be used for information transmission. In 1984 they published the first quantum cryptography protocol now known as "BB84".[5] A further advance in theoretical quantum cryptography took place in 1991 when Ekert proposed[6] that Einstein-Podolsky-Rosen (EPR) entangled two-particle states could be used to implement a quantum cryptography protocol whose security was based on Bell's inequalities.[7] Also in 1991, Bennett and collaborators demonstrated that QKD was potentially practical by constructing a working prototype system for the BB84 protocol, using polarised photons.[8]

In 1992 Bennett published a "minimal" QKD scheme ("B92") and proposed that it could be implemented using single-photon interference with photons propagating for long distances over optical fibers.[9] Since then, other QKD protocols have been published[10] and experimental groups in the UK,[11] Switzerland[12] and the USA[13] have developed optical fiber-based prototype QKD systems. The aim of these experiments has been to show the conceptual feasibility of QKD, rather than to produce the definitive system, or to address a particular cryptographic application. Thus, we can expect that the experiences with the current generation of systems will lead to improvements towards demonstrating the practical feasibility of QKD as well as a definition of the applications where it could be used.

The remainder of this paper is organised as follows. In Sections 2 and 3 we introduce some important ideas from cryptography to explain the significance of QKD. In Section 4 we shall describe the B92 QKD protocol in detail, and then in Section 5 we shall illustrate the immunity of QKD to eavesdropping. In Section 6 we describe some of the practical issues that arise in implementing QKD and in Sections 7 and 8 we shall give some details about the QKD prototype that we have developed at Los Alamos. Finally, in Section 9 we shall present some conclusions and discuss the future possibilities for QKD development.

## 2. Cryptography

To explain the significance of quantum cryptography it is necessary to describe some of the important features (and perils) of cryptography in general. These points can be illustrated with one of the most famous literary examples of a cipher: Sir Arthur Conan Doyle's "The Adventure of the Dancing Men."[1] In this story, Elsie, the American wife of an English gentleman, Hilton Cubbitt, is terrorised by the appearance of chalked stick-figures outside her house. (See Figure 1.) Sherlock Holmes is called in and quickly realises that the figures are not the scribblings of children, but rather are a form of cryptography, in which each letter of the alphabet has been substituted with a stick figure, known only to the sender (Abe Slaney, "the most dangerous crook in Chicago") and the intended recipient, Elsie. This cryptosystem yields to the cryptanalytical powers of Sherlock Holmes, who breaks the cipher after collecting only 62 characters, by observing the relative frequencies of the different characters, identifying the most frequent with the letter "E" and using intuition.[2] With this information the master detective is able to compose his own cryptogram summoning Abe Slaney to Elsie's house. The criminal, believing that only Elsie could have composed a "Dancing Men" message, is promptly arrested by the police on his arrival.

This story illustrates several important cryptographic issues. It shows that the two important and distinct aspects of cryptography (secrecy of communications and authentication) can both be accomplished if sender and recipient share a secret to start with (in this case the letter-equivalent of each "Dancing-Man" character). Secondly, this story shows (one of) the main assumption(s) underlying cryptography: no matter how difficult it might appear to be, we must assume that any cryptogram is susceptible to passive interception. Finally, this story shows the complacency that is characteristic of the history of cryptography: Abe believed that his cipher was "unbreakable"

The "Dancing Men" type of cipher, technically known as a monoalphabetic substitution cipher, is clearly unsuitable for all except the most rudimentary applications, which raises the question: are there shared secrets that are more secure? By "more secure" we mean that the encryption algorithm and shared secret must be capable of ensuring the secrecy of communications for the useful lifetime of the information being transmitted. In practice we can make a distinction between "tactical" and "strategic" security. For example, if the information is tactical military information this time might only be a few minutes or hours, because over longer periods the tactical situation may have changed sufficiently to render the information valueless. On the other hand, there are communications such as diplomatic ones, which must "never" fall to an eavesdropper's cryptanalytical attacks, no matter how much computing power is brought to bear on the problem either now or in the future.[14, 15] Indeed, there is one cipher based on a shared secret which is provably unbreakable, and we shall describe this for illustrative purposes in the next section.

## 3. Key material, key distribution and the "one-time pad"

In modern "secret key," or "symmetric" cryptosystems, the general nature of the encryption algorithm, $E$, by which a plaintext message, $P$, is rendered into a cryptogram, $C$, can be publicly known, because in any particular communication it depends on a parameter, known as a "key", $K$, which is a secret shared only by the sender (known generically as "Alice")[16] and intended recipient (known as "Bob"). Indeed, a second important assumption of cryptography, known as Kirchoff's precept, is that the secrecy must reside entirely in the key and not the algorithm.[2,17] Thus, Alice generates the cryptogram

$$C = E_K(P) \quad , \tag{1}$$

and sends it to Bob, who decrypts it

$$P = E_K^{-1}(C) \quad , \tag{2}$$

recovering the plaintext $P$. The transmission of the cryptogram takes place under the nose of an eavesdropper ("Eve") who clearly must not be able to acquire the key, $K$, or she will be able to read Alice and Bob's communications.

Many different cryptosystems have been constructed around this secret key concept (e.g. the popular DES system uses a 56-bit key[15]), but one, invented in 1917,[18] is provably secure (unbreakable). This is the so-called "one-time pad" in which Alice and Bob possess a quantity of secret key material comprising (truly) random characters (bits, digits or letters) that is as large as the message to be transmitted.[19] In this system, if Alice has a plaintext message, $P$, composed of a character sequence $\{p_1, \dots, p_n\}$ (the $p$s will be bits, digits or letters) she uses her key, $K = \{k_1, \dots, k_n\}$ to produce the cryptogram $C = \{c_1, \dots, c_n\}$, where

$$c_i = p_i + k_i \pmod{N} \quad , \tag{3}$$

using modular arithmetic in the base, $N$, of the message characters ($N = 2$ for bits, 10 for digits, and 26 for letters, which can be given a numerical value in the range 0 - 25 corresponding to their order in the alphabet).

When Bob receives the cryptogram, $C$, he subtracts his key from it, again using modular arithmetic, to recover the plaintext, $P$. There are three advantages to using modular arithmetic: speed; left-to-right encipherment; and isolation of errors to a single character. In 1949 Shannon proved using information theory[20, 19] that this cryptosystem is secure, provided the key material is truly random, and it is used once and only once. The essence of the security is that, for an eavesdropper, any given decryption is as likely as any other because of the randomness of the key.

The one-time pad system was used for German and Soviet diplomatic communications before and during WWII and by Communist spies during the Cold War. It acquires its name from the practice of printing the key in the form of paper pads, each sheet of which would be torn out and destroyed after being used just once. A photocopy of a German diplomatic one-time pad from the World War II era is shown in Figure 2.

However, if the one-time pad is truly unbreakable, why is it not used exclusively? The answer to this question involves the key generation, distribution and management problems (which also apply to any other secret key cryptosystem). First of all, Alice must generate some truly random numbers, which is not as easy as it might seem. For instance, a simple pseudo-random number generator on a computer would not give a secure key because it will always produce the same sequence for a given "seed" value. To avoid being in this "state of sin"[21] a physical "noise" source could be used to generate the key material. A fictional example of such a scheme, using atmospheric radio noise is described in Tom Clancy's novel "The Sum of all Fears."[22]

Once Alice has generated enough key material to encrypt any anticipated communications she must arrange for Bob to receive a copy of the key without Eve being able to obtain even partial knowledge of it. If Alice and Bob are able to meet beforehand they can accomplish this key distribution in secrecy. But if they are unable to meet, and they share no secret key material beforehand, Alice cannot simply transmit the key material to Bob because by the main assumption of cryptography this transmission is susceptible to passive eavesdropping, which would allow Eve to also acquire the key. Indeed, the only secure method of transmitting the one-time pad would require another one-time pad to encrypt it! (It can be proved that if Alice and Bob possess no secret key material initially they cannot establish a certifiably secure shared key.[23, 8]) The conventional approach to this key distribution problem is for Alice and Bob to establish a secure channel, relying on "physical security" which in reality can make it "difficult" but not impossible for third parties (Eve) to acquire key information, and they must store the key material securely until it is to be used. (See Figure 3.)

The necessity for generating, distributing and storing the key material in advance renders the one-time pad system (and other secret key cryptosystems) vulnerable to the "insider threat." Trusted individuals with access to Alice's or Bob's stored key material could copy it and provide it to Eve. Furthermore, the cumbersome logistics of generating the huge quantities of key and transporting it securely make the one-time pad system susceptible to misuse, undermining its security. It is reported[24] that the demands placed on Soviet diplomatic communications at the start of WWII were so great that one-time pads were re-used,[25] allowing cryptanalysts to unmask the Rosenberg spy ring and the atom-spy Klaus Fuchs.[26]

In theory the one-time pad is unbreakable, but in practice, as we have seen, it has been very difficult to use. This is one reason for the popularity of public key cryptography systems which are "difficult," but not impossible, to break, and easy to use.[17] We do not have space here to describe these very interesting systems, except to point out that they provide "conditional" security: Eve's computing power now and in the foreseeable future is assumed to be limited. However, advances in computer algorithms and hardware threaten to undermine the security of these systems. They are potentially vulnerable in the sense that what is "difficult" today may be much less so in the future. A recent example illustrates this point rather well. In 1977 a challenge was made in Scientific American[27] to break a message encrypted using the RSA method.[16, 15, 28] To do so required factoring a 129-digit number into its 64- and 65-digit prime factors, and this was projected to take ~$4 \times 10^{16}$ years (about one million times the age of the universe). However, in 1994 factoring algorithms[29] and networking of computer resources had advanced to the level where the factorisation only took 8 months (4,000 MIPS-years).[30] There is no reason to believe that that this factoring speed could not be further increased. For instance, a recently invented algorithm for the yet-to-be-constructed quantum computer[31] would permit factoring of RSA-129 in a few seconds if it ran at the speed of a desktop PC.[32] Thus, unforeseen improvements in algorithms or computing power would allow Eve to "reach back in time" and break encrypted communications before they had ceased to be of value.

In the light of these observations it would clearly be an important cryptographic breakthrough to have a new communication method over an "open" channel that would allow Alice and Bob to establish a provably secure key which, moreover, would allow them to detect if Eve was monitoring their transmissions. Such a communications system would allow Alice and Bob to establish a secret key at the time it is required for encrypted communication. This key could even be a one-time pad, and so such a scheme would facilitate real-time unbreakable encryption of communications, and avoid the security problems that we have discussed above. Quantum key distribution is precisely such a breakthrough.

## 4. Quantum key distribution

To understand QKD we must first move away from the traditional key distribution metaphor of Alice sending *particular* key data to Bob. Instead, we should have in mind a more symmetrical starting point, in which Alice and Bob initially generate their own, independent random number sets, containing more numbers than they need for the key material that they will ultimately share. Next, they compare these sets of numbers to distil a shared subset, which will become the key material. It is important to appreciate that they do not need to identify *all* of their shared numbers, or even *particular* ones, because the only requirements on the key material are that the numbers should be secret and random. They can attempt to accomplish a secret distillation if Alice prepares a sequence of tokens, one kind for a "0" and a different kind for a "1", and sends a token to Bob for each bit in her set. Bob proceeds through his set bit-by-bit in synchronisation with Alice, and compares Alice's token with his bit, and replies to Alice telling her whether the token is the same as his number (but not the value of his bit). With Bob's information Alice and Bob can identify the bits they have in common. They keep these bits, forming the key, and discard the others. If one of Alice's tokens fails to reach Bob this does not spoil the procedure, because it is only tokens that arrive which are used in the distillation process.

The obvious problem with this procedure is that if the tokens are classical objects they carry the bit values before they are observed by Bob, and so they could be passively monitored by Eve. However, we shall now see that it is possible to generate a secure key if the tokens are quantum objects. We shall describe the B92 QKD protocol in terms of the preparation and measurement of states in a two-dimensional Hilbert space such as that of a spin-1/2 particle. The spin operators $\sigma_1$, $\sigma_2$, $\sigma_3$, obey the algebra

$$\left[\sigma_i, \sigma_j\right] = 2i\varepsilon_{ijk}\sigma_k \quad , \quad i, j, k = 1,2,3 \quad , \tag{4}$$

and we may introduce a basis of states with spin-up ($\left|\uparrow\right\rangle$) or spin-down ($\left|\downarrow\right\rangle$) along the $z$-axis

$$\sigma_3 \begin{cases} \left|\uparrow\right\rangle \\ \left|\downarrow\right\rangle \end{cases} = \begin{cases} +\left|\uparrow\right\rangle \\ -\left|\downarrow\right\rangle \end{cases} \quad , \tag{5}$$

satisfying the orthonormality relations

$$\begin{aligned} \left\langle\uparrow|\uparrow\right\rangle = \left\langle\downarrow|\downarrow\right\rangle &= 1 \\ \left\langle\uparrow|\downarrow\right\rangle &= 0 \end{aligned} \quad . \tag{6}$$

From these states we can also make eigenstates with spin-up or spin-down along the $x$-axis

$$\sigma_1 \begin{cases} \left|\rightarrow\right\rangle \\ \left|\leftarrow\right\rangle \end{cases} = \begin{cases} +\left|\rightarrow\right\rangle \\ -\left|\leftarrow\right\rangle \end{cases} \quad , \tag{7}$$

where $\left|\rightarrow\right\rangle = 2^{-1/2}\left(\left|\uparrow\right\rangle + \left|\downarrow\right\rangle\right)$ and $\left|\leftarrow\right\rangle = 2^{-1/2}\left(\left|\uparrow\right\rangle - \left|\downarrow\right\rangle\right)$.

A (von Neumann) measurement in quantum theory is a projection operator in Hilbert space.[33] For example, a measurement for spin-down along the $z$-axis is represented by the projection operator

$$P_{|\downarrow\rangle} = |\downarrow\rangle\langle\downarrow| \quad , \tag{8}$$

and similarly a measurement for spin-down along the $x$-axis is represented by

$$P_{|\leftarrow\rangle} = |\leftarrow\rangle\langle\leftarrow| \quad , \tag{9}$$

The result of a measurement $P$ on a state $|\psi\rangle$ is given by the "collapse of the wavefunction"

$$|\psi\rangle \rightarrow \begin{cases} \dfrac{P|\psi\rangle}{\|P|\psi\rangle\|} & \text{with probability } \langle\psi|P|\psi\rangle \\[2ex] \dfrac{(1-P)|\psi\rangle}{\|(1-P)|\psi\rangle\|} & \text{with probability } \langle\psi|(1-P)|\psi\rangle \end{cases} \quad , \tag{10}$$

where we shall describe the first outcome as a "pass" and the second as "fail." Here we have defined the norm as

$$\||\phi\rangle\| \equiv |\langle\phi|\phi\rangle|^{1/2} \quad . \tag{11}$$

Thus, the outcome of a measurement in quantum mechanics is, in general, only predictable with some probability.

For the B92 protocol Alice has two non-orthogonal state preparations: $|\uparrow\rangle$ or $|\rightarrow\rangle$; and Bob can make two non-orthogonal measurements: $P_{|\downarrow\rangle}$ or $P_{|\leftarrow\rangle}$. The "pass" probabilities of the various preparation-measurement combinations are given in Table 1.

|  | $|\uparrow\rangle$ | $|\rightarrow\rangle$ |
|---|---|---|
|  |  |  |

| $P_{|\downarrow\rangle}$ | 0 | 0.5 |
|---|---|---|
| $P_{|\leftarrow\rangle}$ | 0.5 | 0 |

**Table 1.** The probabilities that states prepared by Alice (columns) "pass" Bob's measurements (rows).

In the first step of the B92 protocol (see Figure 4) Alice and Bob generate their own independent sets of random numbers. In Step 2 they proceed through their sets bit-by-bit in synchronisation, with Alice preparing a state for each of her bits according to Table 2.

| bit | state |
|---|---|
| 0 | $|\uparrow\rangle$ |
| 1 | $|\rightarrow\rangle$ |

**Table 2**. The states prepared by Alice for each member of her random bit sequence.

Alice sends each state over a "quantum channel" to Bob. (The quantum channel is a transmission medium that isolates the quantum state from interactions with the "environment.") Bob makes a measurement of each state he receives, according to the value of his bit as given by Table 3,

| bit | measurement |
|---|---|
| 0 | $P_{|\leftarrow\rangle}$ |
| 1 | $P_{|\downarrow\rangle}$ |

**Table 3**. Bob's measurements.

and records the result ("pass" = Y, "fail" = N).

Note that Bob will never record a "pass" if his bit is different from Alice's, and that he records a "pass" on 50% of the bits that they have in common. In Figure 4 we see that for the first and fourth bits Alice and Bob had different bit values, so that Bob's result is a definite "fail" in each case. However, for bits 2 and 3, Alice and Bob have the same bit values and the protocol is such that there is a probability of 0.5 that Bob's result is a "pass" in each case. Of course, we cannot predict which one will be a "pass," but the chances are that one will pass and the other fail. In Fig. 4 we choose the "pass" to be the third bit.

In Step 4 (see Fig. 4) Bob sends a copy of his *results* to Alice (but not the measurement that he made on each bit). He may send this information over a conventional (public) channel which may be subject to eavesdropping. Now Alice and Bob retain only those bits for which Bob's result was "Y" and these bits become the shared key material. (In Fig. 4 the third bit becomes the first bit of the shared key.) This procedure distils one shared bit from four initial bits because it only identifies 50% of the bits that Alice and Bob actually have in common. However, this inefficiency is the price that Alice and Bob must pay for secrecy.

## 5. Eavesdropping on B92

We shall now approach the B92 protocol from Eve's perspective to see why it is secure. So, we should set out in detail what it is that Eve wants to accomplish, what knowledge she may be supposed to have, and what she can do to the quantum and public channels. Eve could simply stop any communications between Alice and Bob by disrupting the quantum channel. But the scenario that we should have in mind is that it is much more rewarding for Eve to acquire information about Alice and Bob's communications without being detected than it is to accomplish "denial-of-service." We shall assume (in accordance with the main assumptions of cryptography) that Eve knows the possible state preparations and measurements available to Alice and Bob, but of course no knowledge of their initial random number sets. In addition, Eve can actively monitor the quantum channel, so that she is able to measure and replace states, but we shall only allow her to monitor passively the transmissions on the public channel. (This is a restriction for the simplicity of presentation. We could allow Eve to alter Bob's public transmissions to Alice, but this could be countered with a secure authentication protocol on the public transmissions so that Alice could be sure that they had not been altered in transmission. For instance, if Alice and Bob share a small amount of initial secret key material, which can be replenished from the new key material generated by QKD, Bob can authenticate his public transmissions to Alice.[8, 34]) Alice and Bob have three tools that they can use to detect eavesdropping: they can measure the key generation rate; they can measure the error rate in a portion of the passes; and they can measure the "0"-"1" bias in a portion of the passes.

The first and most obvious aspect of security is that it is impossible for Eve to "tap" the quantum transmissions in the conventional sense: a single quantum cannot be split.[35] But perhaps Eve could use an "amplifier" to clone each of Alice's states, reading the copy and forwarding the original to Bob? However, while an amplifier could faithfully copy orthogonal states, it would require a violation of quantum mechanical linearity for it to be able to make exact clones of both of the non-orthogonal states used by Alice.[36] So, we should consider what happens if Eve makes her own measurements (projections) on Alice's states and sends on the results to Bob. Eve faces an immediate difficulty because the projection operators corresponding to Alice's two state preparations do not commute

$$\left[ P_{|\uparrow\rangle}, P_{|\rightarrow\rangle} \right] \neq 0 \quad . \tag{12}$$

The states are non-orthogonal and so cannot be simultaneous eigenstates of both operators.

A thorough eavesdropping analysis is very lengthy[37] and so we shall restrict ourselves to an illustrative example, in which Eve makes the same projection, $P_{|\uparrow\rangle}$, on every state that Alice transmits, recording the result as "0" if the result is a "pass" and as a "1" if the result is a "fail." Eve then sends the resulting state on to Bob. This tactic allows all of Alice's "0" bits to pass this test, but it also erroneously passes 50% of Alice's "1" bits, giving Eve only a 0.75 probability of correctly identifying a "0". On the other hand, Eve can with certainty identify the 25% of Alice's initial sequence which are the "1" states that fail her test. But, the nature of quantum measurements is such that Eve irreversibly alters all of Alice's "1" states so that 50% of them are $|\uparrow\rangle$-states and the other 50% are $|\downarrow\rangle$-states when they reach Bob. Now, if Bob tests either of these states with his "0"-measurement there will be a 50% probability that the state will pass, in conflict with 0% probability of this happening in the absence of eavesdropping. (There is also a bias introduced into Bob's results: more than 50% of his results are "0"s. Of course, a clever Eve would use a different strategy which did not introduce this asymmetry.)

The net result of Eve's activities is that she has only reliably identified Alice's "1"s, at the expense of introducing a 25% error between Alice's and Bob's key material. Thus, Alice and Bob can sacrifice a portion of their key to test the error rate. If the rate is found to be high they will know that Eve has been listening, and they would not use the key material.

If quantum mechanics was merely a statistical theory as envisioned by Einstein,[38] in which probabilities arose from our inability to produce experimentally dispersion-free ensembles with sharp values for all observables simultaneously, then QKD would not be secure. It is the remarkable feature of quantum mechanics that measurements do not reveal pre-existing values,[39] which is in such strong contrast to classical physics, that ultimately underlies the security of QKD.

## 6. Practical implementations of QKD

Perhaps the most obvious way to implement the QKD quantum channel is with single-photon polarization states, such as the preparation of vertical and right-handed-circular polarizations, and the measurement of horizontal linear and left-handed-circular polarizations. However, another set of single-photon states, which we shall call "phase" states, have the algebraic properties required for quantum cryptography and can be constructed by allowing a photon to impinge on a simple beamsplitter.

We can express the action of a lossless beamsplitter in terms of photon creation and annihilation operators (with frequency and polarization labels suppressed) as a (unitary) transformation from the two "in" modes to the two "out" modes (see Fig. 5a),[40] as

$$a_{out}^{(1)} = 2^{-1/2}\left[a_{in}^{(1)} + ia_{in}^{(2)}\right]$$

$$a_{out}^{(2)} = 2^{-1/2}e^{i\phi_A}\left[a_{in}^{(2)} + ia_{in}^{(1)}\right] \qquad , \tag{13}$$

where we have incorporated an adjustable phase shift, $\phi_A$, in the second output. Here

$$\left[a_{in}^{(1)}, a_{in}^{(2)}\right] = \left[a_{in}^{(1)}, a_{in}^{(2)\dagger}\right] = \left[a_{in}^{(1)\dagger}, a_{in}^{(2)}\right] = 0$$

$$\left[a_{in}^{(1)}, a_{in}^{(1)\dagger}\right] = \left[a_{in}^{(2)}, a_{in}^{(2)\dagger}\right] = 1 \qquad , \tag{14}$$

$$a_{in}^{(1)}|0\rangle = a_{in}^{(2)}|0\rangle = 0$$

where $|0\rangle$ is the "vacuum" (no photon) state, and similarly for the "out" operators.(The factors of "$i$" arise from the phase change on reflection.)

Thus, if we equip Alice with this beamsplitter and she introduces a single photon state at the "in"-port "1"

$$\left|in^{(1)}\right\rangle = a_{in}^{(1)\dagger}|0\rangle \qquad , \tag{15}$$

this amounts to a preparation of the non-orthogonal "out" states

$$\left|\uparrow\right\rangle \equiv 2^{-1/2}\left[a_{out}^{(1)\dagger} + ia_{out}^{(2)\dagger}\right]|0\rangle \quad \text{for} \quad \phi_A = 0$$

$$\text{or} \qquad \qquad . \tag{16}$$

$$\left|\rightarrow\right\rangle \equiv 2^{-1/2}\left[a_{out}^{(1)\dagger} - a_{out}^{(2)\dagger}\right]|0\rangle \quad \text{for} \quad \phi_A = \pi/2$$

Bob may now introduce the above "out" modes into the "in" ports of a second beamsplitter (see Fig. 5b) and add an additional phase, $\phi_B$, to Alice's "out" "1" mode, giving final "out" state destruction operators

$$a_{out}^{(3)} = 2^{-1/2}\left[ie^{i\phi_B}a_{out}^{(1)} + a_{out}^{(2)}\right]$$
$$a_{out}^{(4)} = 2^{-1/2}\left[e^{i\phi_B}a_{out}^{(1)} + ia_{out}^{(2)}\right]$$

(17)

If a detector is placed in Bob's "3" output port (see Fig. 5b), the detection of a photon corresponds to a projection onto the non-orthogonal states

$$|\downarrow\rangle \equiv 2^{-1/2}\left[a_{out}^{(2)\dagger} + ia_{out}^{(1)\dagger}\right]|0\rangle \quad \text{for} \quad \phi_B = \pi$$
$$\text{or}$$
$$|\leftarrow\rangle \equiv 2^{-1/2}\left[ia_{out}^{(1)\dagger} + ia_{out}^{(2)\dagger}\right]|0\rangle \quad \text{for} \quad \phi_B = 3\pi/2$$

(18)

The four states constructed above have the necessary orthogonality properties for B92 QKD. Thus, by combining the two beamsplitters Alice and Bob may construct an interferometric version of QKD (see Fig. 6) where the probability that a photon injected by the laser source is detected is given by

$$P_D = \cos^2\left(\frac{\phi_A - \phi_B}{2}\right) \quad .$$

(19)

Thus, if Alice and Bob use the phase angles $(\phi_A, \phi_B) = (0, 3\pi/2)$ for their "0" bits (respectively) and $(\phi_A, \phi_B) = (\pi/2, \pi)$ for their "1" bits they have an exact representation of B92.

To construct a practical quantum cryptography device using either the polarization or the phase states (or combinations thereof) we must consider the production, propagation and detection of single photons. Surprisingly, it is remarkably difficult to produce experimentally a "single-photon" state because of the Poisson statistics of available light sources. A reasonable approximation is to use a highly attenuated pulsed laser source, with a convenient choice being a probability of 10% that the pulse contains one photon.[8] Of course, this means that

~ 90% of the pulses contain no photon, but this is the price that must be paid (in data rate) for having < 1% of pulses containing two or more photons, which are the ones susceptible to a beamsplitting attack. A more sophisticated light source could be based on the phenomenon of parametric down conversion[41] in which a photon entering a non-linear crystal has a small probability of producing a pair of lower-frequency photons. By triggering on one of the down-converted photons true "single photons" have been produced as the other member of the pair.[42]

At first sight, it appears that free-space QKD would be immensely difficult over any but the shortest distances because of the presence of background photons and the difficulties of directing the photons onto a remote detector. Nevertheless, the first working QKD system was based on polarized photons propagating in a tube over a distance of 30 cm.[8] However, these authors proposed that for longer distance communications optical fibers would be more suitable because the photons would be guided from source to detector. Moreover, because optical fibers are widely used in telecommunications there are commercially available components, possibly allowing a system to be constructed that can perform quantum cryptography over an installed communications system.

Optical fibers, although possessing the good feature of guiding photons from source to detector, bring their own problems that largely determine the operating characteristics of a system. Probably the first question to be answered is: what wavelength should we choose to operate at ? Two factors are relevant to this question. At what wavelengths is single-photon detection possible with non-negligible efficiency? and at what wavelengths do optical fibers have low attenuation?

For photons in the wavelength range of 600 - 800 nm there are commercially available single-photon counting modules based on silicon avalanche photodiodes (APDs), which have high efficiencies (< 90%) and low noise rates ( ~ 50 Hz when cooled to ~ -25 °C). However, the attenuation of (single-mode) optical fibers is quite high in this wavelength range (~ 3 dB/km),[43] which will adversely affect the data rate and the noise rate if we choose to operate in this region. (The loss mechanism is predominantly Rayleigh scattering out of the fiber.)

Conversely, optical fibers have much lower attenuation at 1.3 μm (~ 0.3 dB/km), and lower again at 1.55 μm, but although there are commercially available germanium (Ge) and indium-gallium arsenide (InGaAs) APDs that are sensitive to light at these infrared wavelengths, there are no commercially available single-photon counting modules. Nevertheless, several groups have shown that these devices can detect single photons at 1.3 μm if they are first cooled to reduce noise, and operated in so-called Geiger mode, in which they are biased above breakdown.[44] An incoming photon liberates an electron-hole pair, which with some probability initiates an avalanche current, whose detection signals the arrival of the photon. For our project we decided that the propagation distance advantages of the 1.3-μm wavelength were such that we characterized the performance of several APDs (both Ge and InGaAs) for single-photon detection at this wavelength.

Several parameters are important in characterizing the detector performance: single-photon detection efficiency; intrinsic noise rate (dark counts); and time resolution. We measured absolute detection efficiencies of 10 - 40%, (for InGaAs APDs), but noise rates that are ~ 1,000 times higher than for Si-APD photon counting modules at 800 nm. However, our detectors also have very good time resolutions (~ few 100 ps), which can be utilised to compensate for the higher intrinsic noise rate because of the low dispersion of optical fibers at 1.3 μm. Thus, if a 1.3-μm photon is injected into a fiber in a short wavepacket (300-ps, say) it will emerge for the far end without being significantly stretched out in time and so, because we know that the photon will be expected within a short time window we need only consider the probability of a noise count in this short time interval (~ $5 \times 10^{-6}$ for 50-kHz noise rate).

Once we had established that single 1.3-μm photons can be detected with acceptably low-noise background, we had to decide which photon states are most suited for quantum key distribution: polarization and/or phase? For a polarization scheme we would have to propagate two non-orthogonal polarization states down the fiber. An immediate difficulty is that we need to know the relationship between vertical polarization (say) at the fiber input and the output polarization, because the birefringence introduced by bends in the fiber will, in general, convert a linear polarization input state into an elliptically polarized output. Furthermore, this birefringence means that a given length of fiber will, at any instant, have stable "fast" and "slow" propagation modes which will be orthogonally polarized. The difference in propagation speeds[45] (polarization mode dispersion or PMD) typically amounts to about a $\pi$-phase difference over (a beat length of) 10 cm to 1 m of fiber.[46] So, if we align one of the QKD polarization states with the fast axis (say) then the other non-orthogonal state will be gradually depolarized during propagation once its two component polarizations become separated by more than the coherence length of the light source. Thus, PMD will make a polarization QKD scheme difficult for long propagation distances (> few km). Nevertheless two groups have demonstrated polarization-based QKD over ~ 1 km of fiber. However, we decided to avoid these difficulties of polarization-encoding in-fiber and to use instead a phase-encoding scheme similar to the BT group's in UK,[11] who have already demonstrated QKD over 10 km of fiber in a laboratory. As we shall see we still need to control the photons' polarization in this method, but we avoid the problems of non-orthogonal polarization states.

## 7. Single-photon interference for optical fiber QKD

If we were to simply use optical fibers for each of the interfering paths in Figure 6 we would have a very unstable interferometer for all but the shortest propagation distances. However, a more stable system can be produced by multiplexing both paths onto a single fiber in a design first proposed by Bennett.[9] In this design (see Fig. 7) Alice and Bob have identical, unequal-arm Mach-Zehnder interferometers with a "short" path and a "long" path, with one output port of Alice's interferometer optically coupled to one of the input ports of Bob's. The difference of the light travel times between the long and short paths, $\Delta T$, is much larger than the coherence time of the light source, so there can be no interference within each small interferometer. However, interference can occur within the coupled system.

A photon injected into one of the input ports of Alice's interferometer from the attenuated pulsed laser source ("$L$" in Figure 7) therefore has a 50% probability of entering Bob's interferometer, in a wave packet that is a coherent superposition of two pieces that are separated in time by $\Delta T$, corresponding to an amplitude for it to have taken the "short" path, and a delayed component which took the "long" path. On entering Bob's interferometer each component of the wave packet is again split into a "short" component and a "long" component, so that at each output port there are three "time windows" in which the photon may arrive. The first of these ("prompt") corresponds to the "short-short" propagation amplitude; which is followed after a delay of $\Delta T$ by the "central" component comprising the "short-long" and "long-short" amplitudes; and finally, after a further time $\Delta T$, the "delayed" time window corresponds to the "long-long" amplitude.

There is no interference in the "short-short" or "long-long" amplitudes, so the probability that the photon arrives in either of these time windows in either of Bob's output ports is 1/16 (we assume 50/50 beamsplitters and lossless mirrors). However, because the path-length differences in the two small interferometers are identical (to within the coherence length of the light source) interference does occur in the "central" time window between the "short-long" and "long-short" amplitudes. Indeed, because Alice and Bob can control the path length of their "long" paths with adjustable phases $\phi_A$ or $\phi_B$, respectively, the probability that the photon emerges in the "central" time window at the detector ("$D$" in Figure 7) in the output port shown is

$$P = \frac{1}{8}\left[1 + \cos\left(\phi_A - \phi_B\right)\right] \quad . \tag{20}$$

Note that within a factor of four this expression is identical with the photon arrival probability for the simple interferometric version of B92, and that, of the two interfering paths one ("long-short") is controlled by Alice and the other ("short-long") is controlled by Bob just as in the simple interferometer of Fig. 6. Thus, by sacrificing a factor of four in data rate this time-multiplexed interferometer can be used to implement QKD based on single-photon interference. (The photons "lost" in the prompt and delayed time windows are useful to test for a highly invasive Eve.[9])

In order to be sure that we understood single-photon interference in such a configuration we first constructed a time-multiplexed interferometer, from beamsplitters and mirrors, with ~ 500-ps path differences, and directed highly attenuated 30-ps pulses of 1.3-μm light with a 10-kHz repetition rate into one of the input ports. At one of the output ports light was collected by a microscope objective and directed onto one of our cooled InGaAs APDs, from which a photon time-of-arrival spectrum was accumulated with the mirrors in a fixed position. Then the mirrors were moved to a new position and a new time spectrum recorded. As expected, interference was observed in the central time window with a number of counts corresponding to an average number of photons per pulse of 0.07. Thus interference was seen with only one photon in the device at a time.

We next constructed an optical fiber version of the time-multiplexed interferometer in which each of Alice's and Bob's interferometers are built from two 50/50 fiber couplers instead of beamsplitters. (See Figure 8.). Each coupler has two input legs and two output legs: a photon entering on one leg has a 50% probability to emerge from either of the output legs. No mirrors are required because the output fiber legs from the first coupler convey the photons to the input legs of the second coupler via a long fiber path or a short path ($\Delta T \sim 8.5$ ns). One of the output legs of Alice's interferometer is connected by a long optical fiber path to one of the input legs of Bob's interferometer. Finally, photons emerge from one of the output legs of Bob's interferometer into a fiber pigtailed, cooled InGaAs APD detector.

For our first system we constructed the entire optical path from so-called polarization-maintaining (PM) fiber in order to ensure that the two interfering amplitudes had identical polarizations, and also because the electro-optic phase shifters in each of the "long" paths, which we use to adjust the optical path lengths, only leave specific linear polarizations unchanged. (We also included an air-gap in Alice's "short" path so as to adjust the lengths of the two interfering paths to be equal.) The phase modulators are fiber-pigtailed devices in which photons propagate through a waveguide in a non-linear crystal whose refractive index changes when an electric field is applied. The amount of phase change, $\phi$, for a given voltage, $V$, can be expressed as

$$\phi = \cos^{-1}\left(\frac{V}{V_\pi}\right) \quad , \tag{21}$$

where the voltage, $V_\pi$, required for a $\pi$-phase change is about three times greater for the TE mode than the TM mode. (In our interferometer the photon polarization was therefore aligned for TM-mode propagation. Phase modulation rates of up to several GHz are possible with these devices.) For longer propagation distances, where we will have to use ordinary single-mode fiber, we will use a polarization controller[47] in Bob's interferometer to rotate the polarization back to the linear state required for his phase modulator.

A linearly polarized "single-photon" is generated by applying a 300-ps electrical pulse to a low-power, fiber-pigtailed semiconductor laser whose output in then attenuated before coupling into the interferometer. The electrical pulse is also the "start" signal for a time-interval analyzer and applies the pulsed-bias gate signal to the detector after a delay corresponding to the light transit time through the system. The detector avalanche acts as the "stop" signal. Figure 9 shows a time spectrum of photon arrival times. The 8.5-ns separation of the different paths is clearly visible, as is the 300-ps width of the laser pulse. The unequal height of the "short-short" and "long-long" peaks is due to attenuation in the phase modulators.

To turn this optical system into a QKD device we have placed it under the control of two personal computers: one to control the overall timing and to set Alice's phases, and the other to set Bob's phases and record his results (see Figure 10). The computers are linked together over Ethernet, forming the public channel, in order to initialize their activities and to perform the results-transfer step of the QKD. A key sending procedure starts with Alice's and Bob's computers first generating independent sets of 1,024 random bits. (We used pseudorandom number generation for this experiment.) Each of these sets of bits is loaded into the memory of a digital-to-analogue converter (DAC). When this is accomplished, key sending starts under the control of Alice's master timer. In our first experiment this clock ran at a rate of 100 kHz, but used ten "ticks" of the clock to set one bit. During this time: the contents of the one memory location in each DAC were used to set voltages on the phase modulators corresponding to the B92 protocol; the laser was pulsed once and the detector gated open; and the output of the detector in the central time window recorded to Bob's computer. The procedure was then repeated with the next bit, and so on, until all 1,024 bits were used up. At the end of this procedure, Bob's computer had a file recording the bit number (1 - 1,024), the bit value ("0" or "1") and whether the detector fired or not ("hit" or "miss"). Then, Alice's computer received over the Ethernet link a file from Bob of the "hit" or "miss" status for each bit number. With this information Alice and Bob retain only the bits corresponding to "hits," which become the key material. More sets of 1,024 bits are generated and sent until a long enough key is built up to encrypt the message that is to be sent.

We have carried out key generation with this system over a short distance and are now rebuilding the system to improve its stability and to go to longer distances. We are able to have fiber propagation distances of up to 1 km within our laboratory, and in the future we hope to explore the practical feasibility of our QKD system over a distance of 15 km of installed fiber that links two of the Technical Areas at Los Alamos.

In our system it will be possible to type in a short message on Alice's computer, which will then convert it into ASCII code and generate as many key bits as there are bits in the code. The message will be encrypted by adding the key to the ASCII codes, the characters corresponding to the new ASCII codes will be displayed on Alice's computer, and sent over the Ethernet to Bob's computer where they are displayed again. Bob's computer will then subtract its key to recover the original message.

## 8. Data rate and error rate

Several factors make the key generation rate of any QKD system considerably slower than the laser pulse rate. Firstly, the "single-photon" requirement introduces a factor of ten reduction in rate because 90% of the laser pulses contain no photon. Then there are attenuation losses during propagation, which would amount to about a factor of four over 10 km of optical fiber. The QKD procedure itself has an intrinsic inefficiency of only identifying one shared bit from four initial bits, which is reduced by a further factor of four in our scheme, resulting in an additional factor of sixteen reduction in key rate. Finally, there is the detector efficiency to be included, which in our case was only about 20%. Overall, the key generation rate with these parameters is about 1/3,000 of the laser pulse rate, and would be reduced further over greater propagation distances because of the additional fiber attenuation. However, the useful distance is limited not so much by the key rate as by the bit error rate (BER) which must be less than some threshold in order that eavesdropping can be detected.

One of the contributing factors to the error rate is the interferometric visibility (the contrast between bright and dark fringes). For example, with a 99.5% visibility there is a small probability that a photon would trigger Bob's detector even if his bit value is different from Alice's, introducing a 0.5% BER into the key generation. However, for long propagation distances, a more important error source is from detector noise. There is a fixed probability of a noise count occurring in Bob's detector that is independent of propagation distance, but because the number of photons arriving decreases with increasing distance, the noise contribution to BER actually increases with distance. Thus, for a given detector efficiency there is a maximum useful propagation distance at which the limiting BER for detecting eavesdropping is reached. There will be a trade-off between key-rate, which can be increased by increasing detector efficiency, and BER which also increases with efficiency. A major improvement in BER for a given propagation distance could be accomplished with a genuine "single-photon" source such as the one described earlier, because one could then avoid the factor of ten reduction in key rate required with the highly attenuated laser light of our prototype.

Another way to increase the key rate would be to simply increase the laser pulse rate, but we would quickly run into a problem of increased detector noise arising from "after-pulsing." This is a phenomenon whereby charges that flowed during a detector avalanche become "trapped" in the device until it is gated open for the next pulse, when they are released, leading to a spurious detection. The trapping phenomenon has an exponential decay with a time constant of a few microseconds for our InGaAs detectors and so imposes an upper limit on our detection rates of about 100 kHz.

Because there will inevitably be errors in Alice's and Bob's keys there is a need for error correction. In our prototype this is accomplished by a simple block-parity check, to identify blocks that contain an error, followed by discarding one bit from each block to ensure that an eavesdropper cannot acquire some key information. An additional procedure called "privacy amplification" can be used with public communications to reduce any partial key information obtained by Eve to an arbitrarily low level.[48]

## 9. Summary

The conceptual feasibility of quantum key distribution has now been demonstrated in the laboratory by four groups including ourselves. It is clear that propagation distances of ~ 50 km with key generation rates of a few kHz are already possible. The next step is to investigate the practical feasibility of the technology over installed optical fiber links outside the laboratory. Several developments will be desirable. We have already discussed the advantages to be gained from a true "single-photon" light source. The limiting factor in key rate and propagation distance at present is from detector noise, so detector development work on dedicated high-efficiency, low-noise single-photon counting modules for the 1.3-μm and 1.55-μm wavelengths would be very useful. In addition, the current QKD designs try to control two dimensions of the photon's four-dimensional Hilbert space (polarization in our case) while encoding the key information in the other two dimensions (phase in our design). However, it may be possible to produce better designs, utilizing all four dimensions, that have superior stability properties.

QKD has two attractive features for symmetric cryptosystems. It offers the ultimate security of the inviolability of a law of Nature for key distribution, and it introduces an "ease of use" aspect: key material can be generated when it is required, avoiding the cumbersome, time-consuming security measures of conventional key distribution. QKD could be used to generate any shared key, from 56-bit DES keys, all the way to one-time pads for unbreakable encryption. So, we should ask: "Where could optical-fiber based QKD be used?" An obvious possibility is for absolutely secure communications between "secure islands" over "open" fiber links. For example, one could imagine that it might be used between different US government agencies within the Washington DC area, or in the UK between GCHQ and Whitehall. However, because amplifiers would destroy the quantum coherence used in QKD, transmission distances much greater than 100 km would only be possible through secure "repeater" stations, where key material would be generated for retransmission of a message to the next station. So, for long distances the feasibility of free-space QKD needs to be studied.

QKD is the first practical application of the foundations of quantum mechanics, and as such it attests to the value of basic science research. However, if QKD is to ever be used in practice its security must be certified, and so we should examine with great thoroughness the aspects of quantum mechanics on which its security is based. In order to validate these security concepts it may even be necessary to perform new experiments on the foundations of quantum mechanics. Thus, we should expect there to be considerable feedback from QKD into basic physics, perhaps leading to a new perspective on the foundations of quantum mechanics, which will be more "practical" than "philosophical." In any event, we can look forward to an exciting future for QKD with many possibilities for future theoretical, experimental and applied physics research.

## Acknowledgements

RJH wishes to thank the Dean and Students of Christ Church, Oxford, England for their hospitality while he was the Dr. Lee Visiting Fellow. Helpful conversations with S. Barnett, H. Brown, M. Edwards, A. Ekert, J. Franson, N. Gisin, N. S. P. King, B. Marshall, P. Milonni, J. D. Murley, J. Paton, J. Perrizo, J. Rarity, C. Simmons, P. Tapster, and P. Townsend are gratefully acknowledged. RJH also wishes to thank Mr. L. Sharp and the National Cryptological Museum for their help and permission to use Figure 2. The research described in this paper was performed with US government funding.

**Figure Captions**

Figure 1. Sherlock Holmes' "Dancing Men," which spell out the message "AM HERE ABE SLANEY."

Figure 2. A page from a German World War II era one-time pad (left-hand side of page).

Figure 3. Key distribution in conventional cryptography.

Figure 4. The four steps in B92 QKD. See text for details.

Figure 5a. Relationship between photon destruction operators at Alice's beamsplitter ("BS") used to prepare QKD states.

Figure 5b. Relationship between photon destruction operators at Bob's beamsplitter ("BS") used to measure QKD states.

Figure 6. Realization of B92 QKD using single-photon interference in a Mach-Zehnder interferometer built from beamsplitters ("BS") and mirrors ("M"). Alice has a source of single photons, one for each initial bit, and controls the lower optical path length. Bob has a single-photon detector and controls the upper optical path length according to the protocol described in the text. The graph at right shows the probability that a photon is detected when Alice and Bob have different bits (zero) and when they have the same bits (0.5). The phase angles shown for the first four photons correspond to the bit values of the example in Figure 4.

Figure 7. A time-multiplexed version of the interferometer of Figure 6 constructed from two, smaller, unequal arm Mach-Zehnder interferometers. A single photon is produced in a short wavepacket at the laser source "L" at left and detected at the detector "D" at right. The photon has a 50% probability to exit the first interferometer through the port that is linked to the second interferometer. It will be in a coherent superposition of an amplitude to have propagated via the "short" path through the first interferometer, and a temporally delayed amplitude corresponding to "long" path propagation. Each component is again split at the second interferometer, leading to three time-separated windows in which the photon may arrive at the detector. Single-photon interference occurs in the central time window between the "short"-"long" and "long"-"short" propagation amplitudes, so that the photon detection probability in this time window varies with the difference of the adjustable phases $\phi_A$ and $\phi_B$ in each interferometer.

Figure 8. Schematic representation of the optical system for the Los Alamos QKD prototype. The device is an all-fiber realization of the interferometer shown in Figure 7.

Figure 9. Time-of-arrival spectrum of single photons propagating through the fiber interferometer of Figure 8. The leftmost peak corresponds to photons that traveled by both "short" paths and the rightmost peak to those that took both "long" paths. The number of photons in the central peak varies with the difference of the phases in each interferometer because of interference between the "short"-"long" and "long"-"short" amplitudes.

Figure 10. Schematic representation of the complete QKD system. Twin computer control systems prepare and measure single photons, produced at the laser "L" and detected at the detector "D," using the optical system (quantum channel) of Figure 8. The reconciliation of the results and key distillation occurs over an Ethernet link between the two computers (public channel). The distillation of one key bit from four initial bits is shown corresponding to the example from Figure 4.

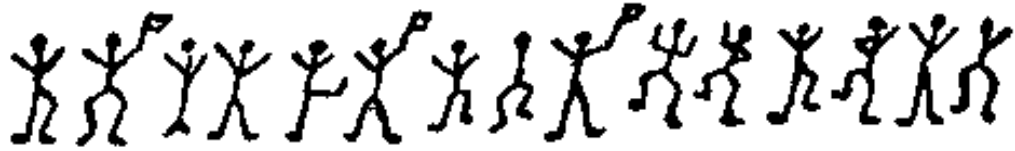

Figure 1. Sherlock Holmes' "Dancing  Men" which spell out the message "AM HERE ABE SLANEY."

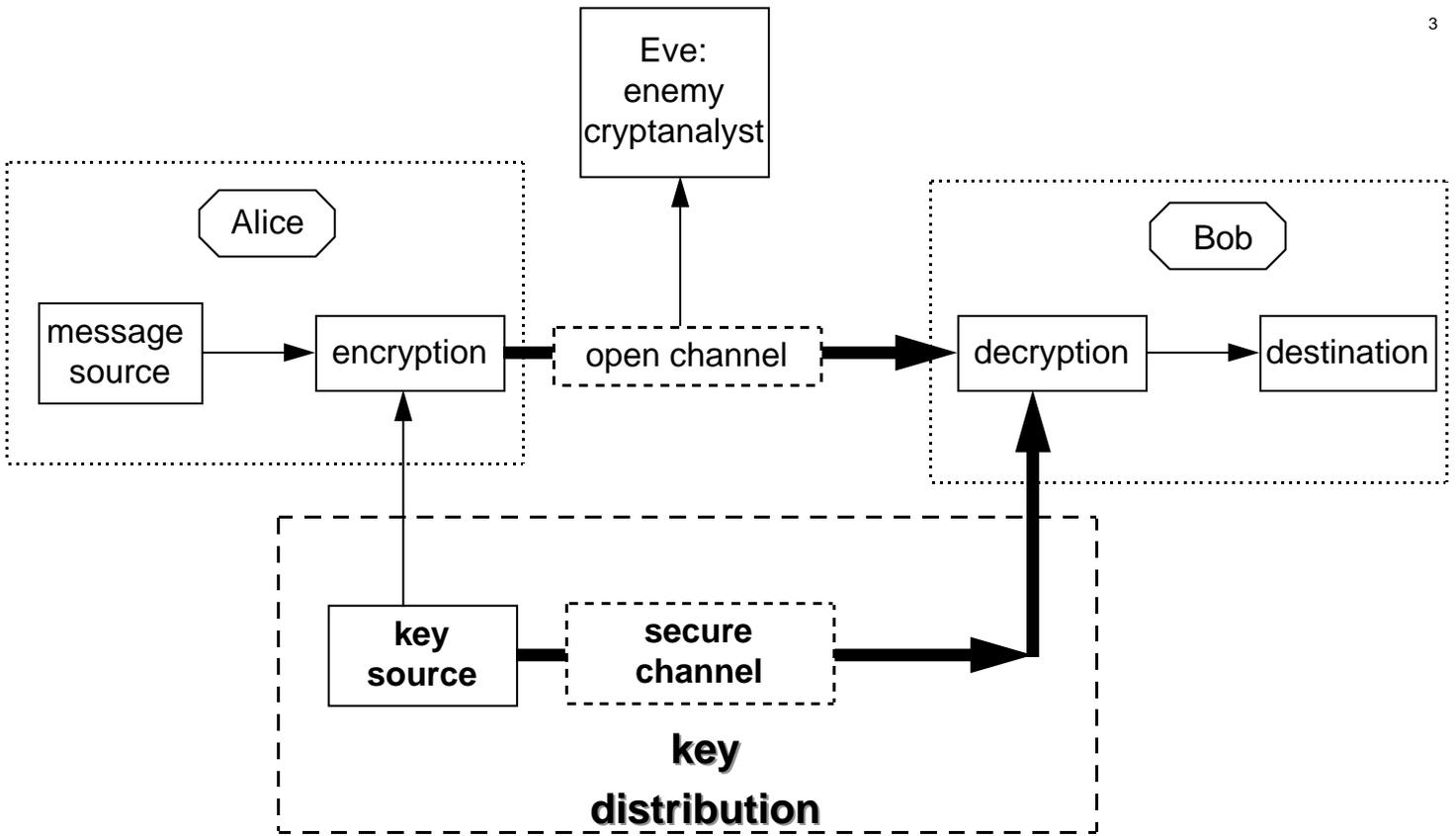

Figure 3. Key distribution in conventional cryptography.

**Step 1**: Alice & Bob generate
**independent** random bit sets

| Alice | 1 | 0 | 1 | 0 | ... |
|-------|---|---|---|---|-----|

| Bob | 0 | 0 | 1 | 1 | ... |
|-----|---|---|---|---|-----|

**Step 2**: Alice sends Bob

"0" $\leftrightarrow$ | $\rangle$    or    "1"   | $\rangle$

**Bob measures (projects)**

**Step 4**: Alice and Bob retain only the "Y"
bits: perfectly correlated subset

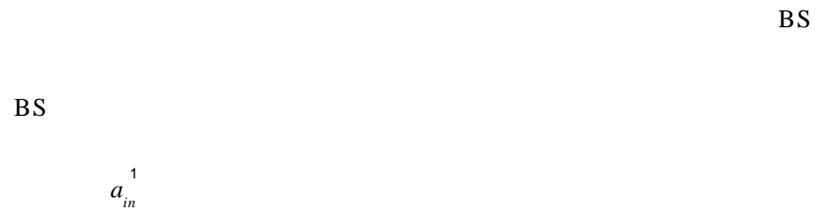

Figure 5a. Relationship between photon destruction operators at Alice's beamsplitter ("BS") used to prepare QKD states.

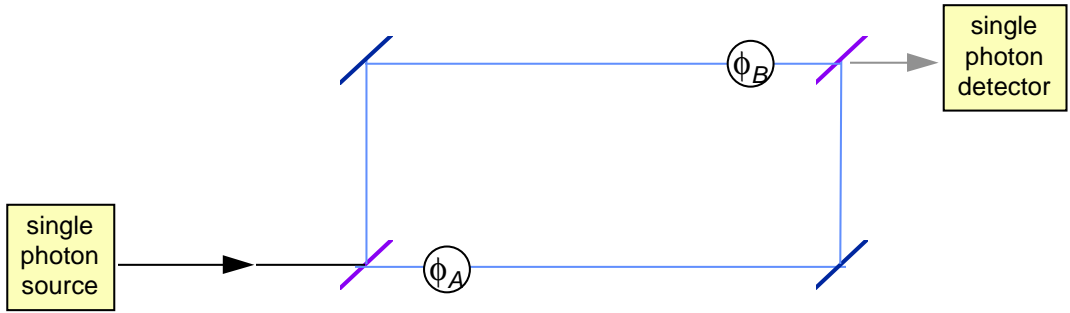

$$P_D = \frac{A - B}{}$$

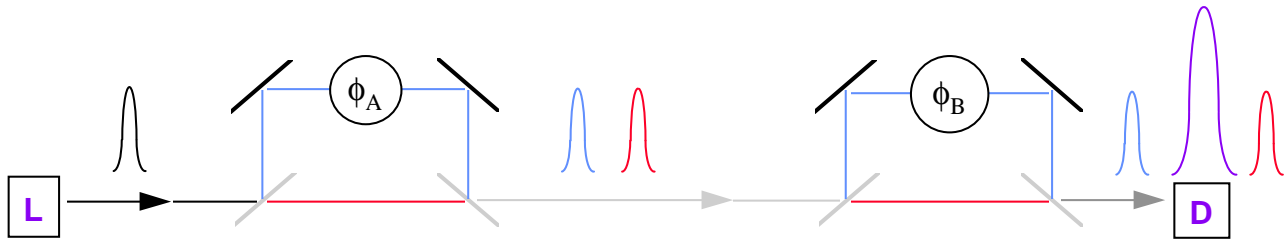

Figure 7. A time-multiplexed version of the interferometer of Figure 7 constructed from two, smaller, unequal arm Mach-Zehnder interferometers. A single photon is produced in a short wavepacket at the laser source "L" at left and detected at the detector "D" at right. The photon has a 50% probability to exit the first interferometer through the port that is linked to the second interferometer. It will be in a coherent superposition of an amplitude to have propagated via the "short" path through the first interferometer, and a temporally delayed amplitude corresponding to "long" path propagation. Each component is again split at the second interferometer, leading to three time-separated windows in which the photon may arrive at the detector. Single-photon interference occurs in the central time window between the "short"-"long" and "long"-"short" propagation amplitudes, so that the photon detection probability in this time window varies with the difference of the adjustable phases $\phi_A$ and $\phi_B$ in each interferometer.

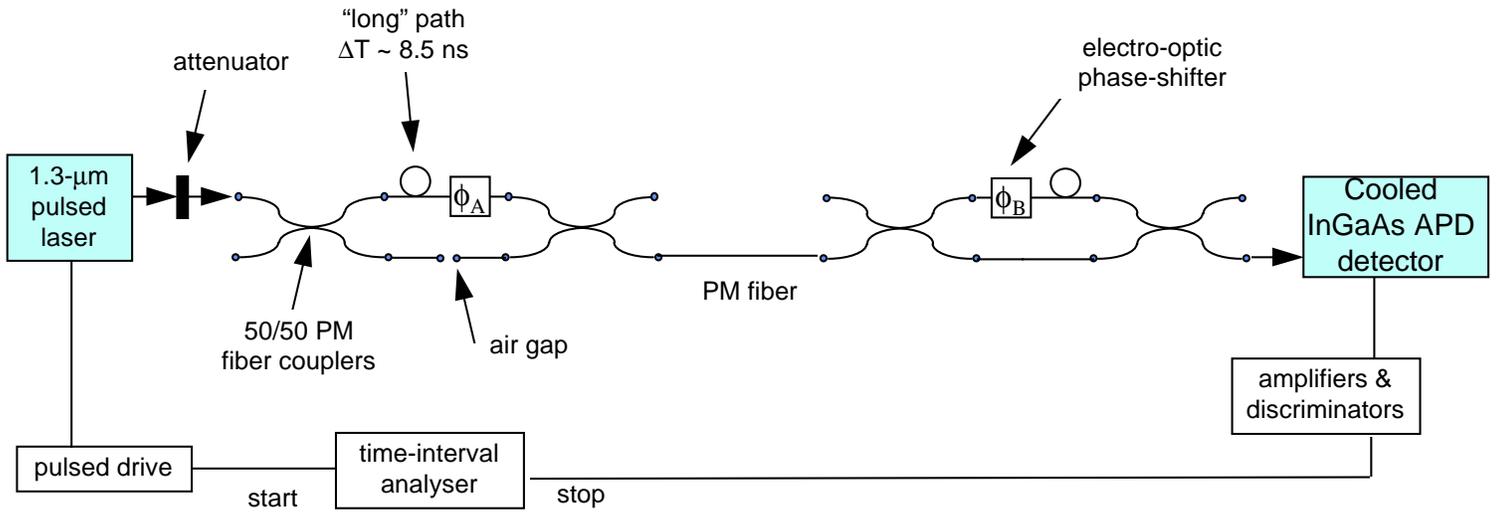

Figure 8. Schematic representation of the optical system for the Los Alamos QKD prototype. The device is an all-fiber realization of the interferometer shown in Figure 7.

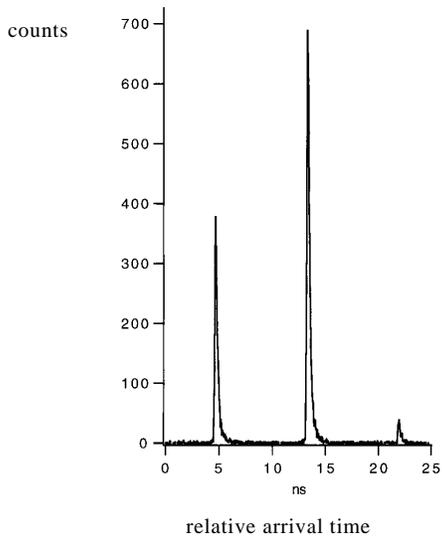

counts

relative arrival time

Figure 9. Time-of-arrival spectrum of single photons propagating through the fiber interferometer of Figure 8. The leftmost peak corresponds to photons that travelled by both "short" paths and the rightmost peak to those that took both "long" paths. The number of photons in the central peak varies with the difference of the phases in each interferometer because of interference between the "short"-"long" and "long"-"short" amplitudes.

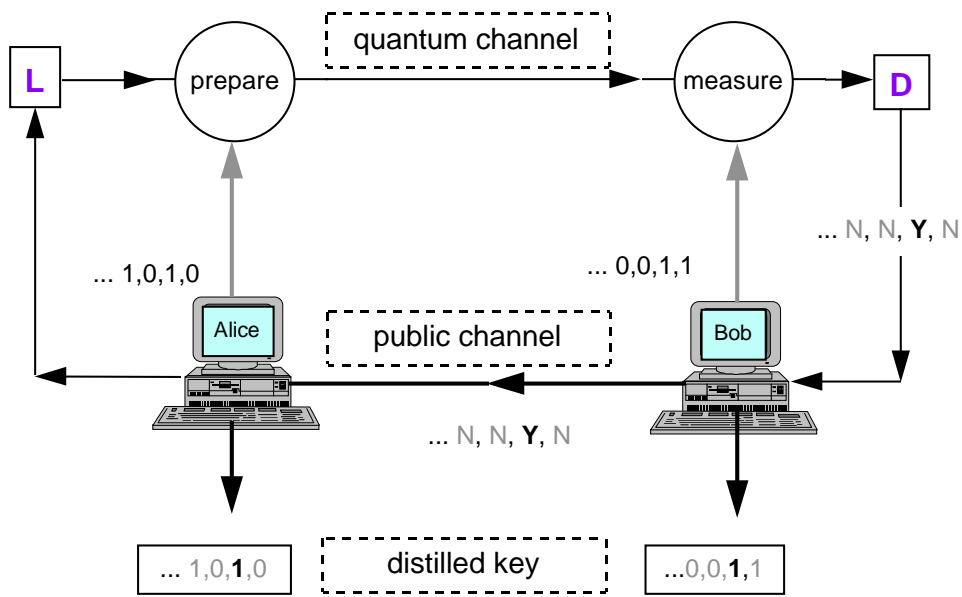

Figure 10. Schematic representation of the complete QKD system. Twin computer control systems prepare and measure single photons, produced at the laser "L" and detected at the detector "D," using the optical system (quantum channel) of Figure 8. The reconciliation of the results and key distillation occurs over an Ethernet link between the two computers (public channel). The distillation of one key bit from four initial bits is shown corresponding to the example from Figure 4.